\documentclass[aps,physrev,twocolumn,amsmath,amssymb,superscriptaddress,10pt,nobibnotes]{revtex4-2}
\usepackage{preamble}

\begin{document}
	

\title{Temporal shaping of wave fields for optimally precise measurements in scattering environments}
\author{Dorian Bouchet}
\altaffiliation{\href{mailto:dorian.bouchet@univ-grenoble-alpes.fr}{dorian.bouchet@univ-grenoble-alpes.fr}}
\affiliation{Universit\'e Grenoble Alpes, CNRS, LIPhy, 38000 Grenoble, France}
\author{Emmanuel Bossy}
\affiliation{Universit\'e Grenoble Alpes, CNRS, LIPhy, 38000 Grenoble, France}


\begin{abstract}
A wave propagating through a scattering medium typically yields a complex temporal field distribution. Over the years, a number of procedures have emerged to shape the temporal profile of the field in order to temporally focus its energy on a receiver. By analogy, we theoretically and experimentally demonstrate here how to maximize the total Fisher information transmitted to a receiver, and how to focus the Fisher information at any given time. This enables one to estimate small variations in the value of any physical observable with optimal precision from noisy measurements, as experimentally illustrated using acoustic waves in the ultrasound regime. By yielding the ultimate precision limit achievable from time-resolved measurements performed in arbitrarily complex media, our approach sets a general benchmark for many applications such as structural health monitoring and biomedical imaging. 
\end{abstract}

\maketitle


\section{Introduction}

Many innovative techniques for focusing and imaging inside scattering media are based on the possibility to control the propagation of waves in space and time from the far field~\cite{mosk_controlling_2012}. Using acoustic waves, pioneer experiments have demonstrated how to focus the field energy within scattering media~\cite{prada_decomposition_1996,kuperman_phase_1998}, to experimentally measure propagation operators and scattering matrices~\cite{aubry_random_2009,gerardin_full_2014}, and to localize acoustic sources in complex environments~\cite{borcea_imaging_2002,ing_solid_2005,qiu_time_2011,ciampa_impact_2012}. This laid the foundations to the development of wavefront shaping protocols at optical frequencies~\cite{rotter_light_2017,gigan_roadmap_2022}, which have been since then widely applied to maximize the energy delivered inside or behind complex scattering media~\cite{vellekoop_focusing_2007,vellekoop_universal_2008,popoff_measuring_2010,kim_maximal_2012,cheng_focusing_2014,sarma_control_2016,bender_depth-targeted_2022}, not only in space but also in time~\cite{aulbach_control_2011,katz_focusing_2011,mccabe_spatio-temporal_2011,mounaix_deterministic_2016,jeong_focusing_2018,xiong_long-range_2019,devaud_temporal_2022}. 

In the recent years, spontaneous motions and intrinsic permittivity variations of a target scatterer have emerged as feedback mechanisms to enhance the energy deposited at the position of the target~\cite{ma_time-reversed_2014,zhou_focusing_2014,ruan_focusing_2017}. It was then realized that the fields generated by small perturbations of any observable $\theta$ characterizing the target can be used to maximize the conjugate quantity to $\theta$ \cite{ambichl_focusing_2017,horodynski_optimal_2020,del_hougne_coherent_2021}. Thus, depending on which observable is perturbed, one can choose to apply the strongest possible force, pressure or torque upon this target, resulting in the formation of completely different fields that do not necessarily maximize the wave energy at the target position. Equivalently, it was shown that a similar procedure allows one to determine the input wave that maximizes the Fisher information carried by the output field~\cite{bouchet_influence_2020,bouchet_maximum_2021}, i.e., to optimize the precision at which it is possible to estimate the value of $\theta$ from noisy measurements of the output field. However, such wavefront shaping methods were developed so far only to optimize the spatial distribution of monochromatic waves, thereby discarding the temporal dimension of the field.

In this article, we theoretically and experimentally demonstrate how to temporally shape input wave fields in order to maximize the Fisher information carried by measured output fields for any given observable $\theta$, regardless of the complexity of the scattering medium in which the waves propagate. To this end, we show how the concept of the Fisher information operator, which was originally introduced for spatially-resolved measurements in the monochromatic regime~\cite{bouchet_maximum_2021}, can be extended to maximize the Fisher information in the time domain. Using this formalism, we introduce fundamental insights derived for static systems that are time invariant, including a deep connection to time-reversal experiments. Finally, we experimentally demonstrate our findings with ultrasound waves, showing that temporally-shaped input fields can not only maximize the total Fisher information carried by the output field, but can also focus the Fisher information at any given time. Our optimal procedure, which is broadly applicable to acoustic and electromagnetic waves, sets a general benchmark for metrology and imaging applications~\cite{barrett_foundations_2003}.


\section{Fisher information in time-resolved measurements}

\begin{figure*}[t]
	\begin{center}
	\begin{minipage}{.495\linewidth}
		\includegraphics[width=\linewidth]{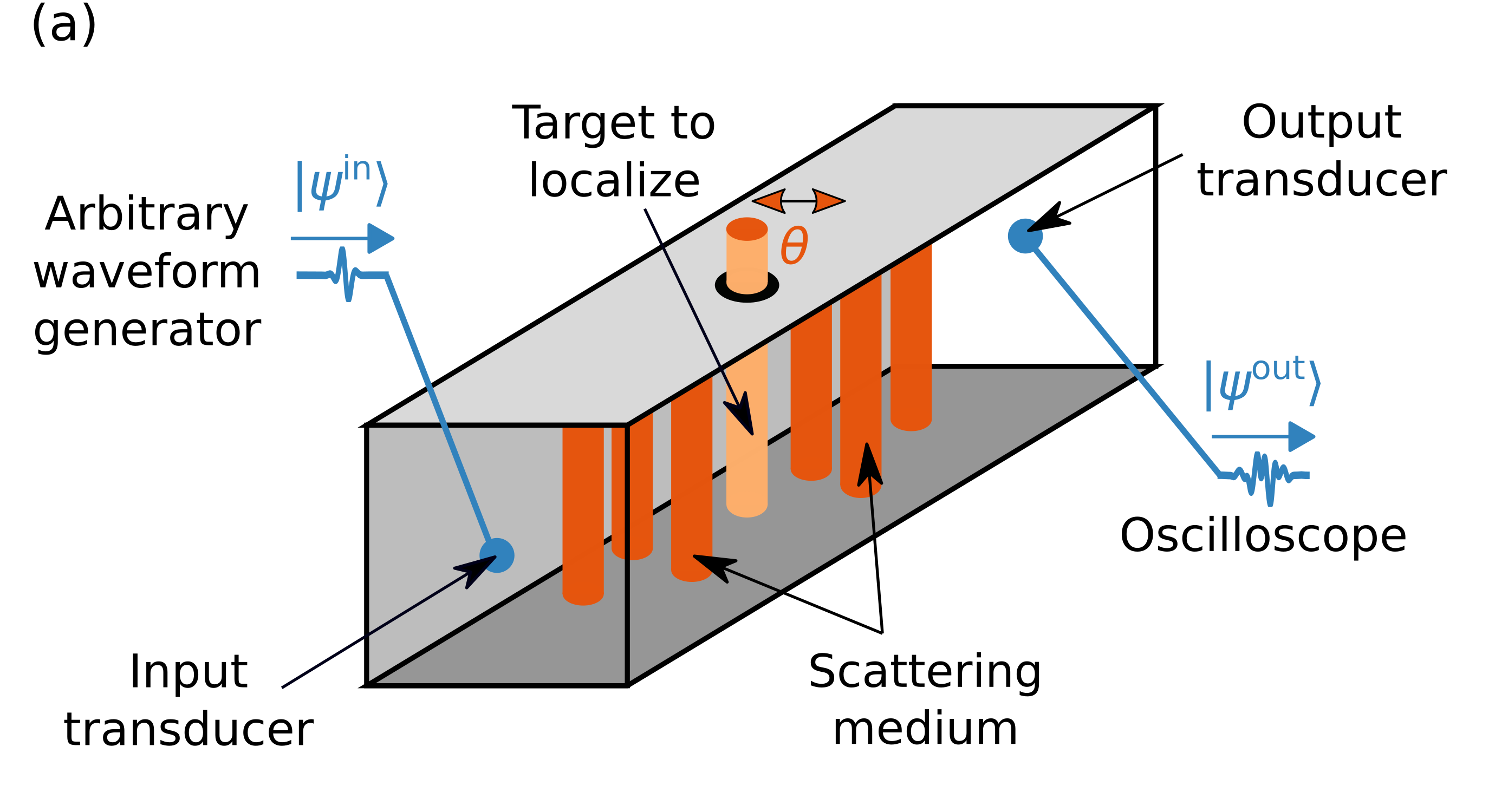}
	\end{minipage}
	\begin{minipage}{.495\linewidth}
		\includegraphics[width=8.6cm]{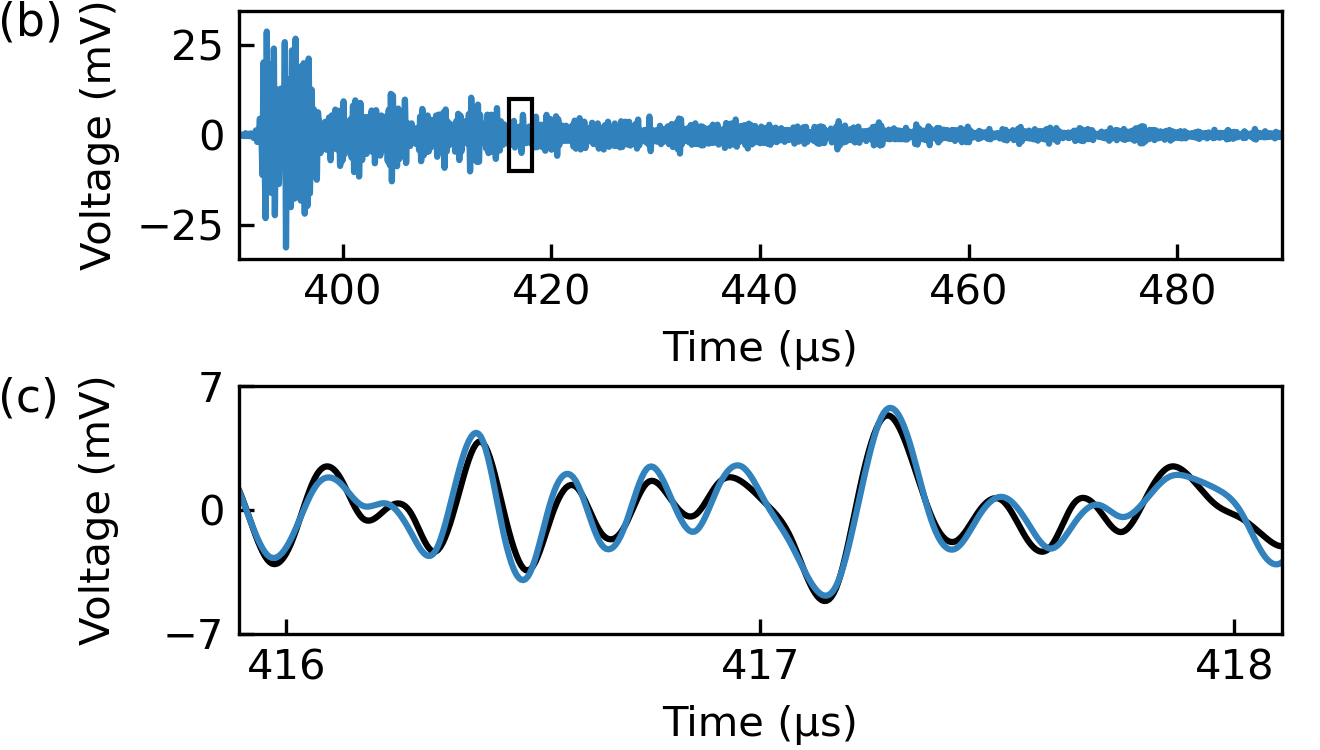}
	\end{minipage}
	\end{center}
	\caption{(a)~Representation of the experiment. A waveform generator connected to a transducer is used to temporally shape the input field. This field propagates within a complex scattering system and reaches an output transducer connected to an oscilloscope. From the measured output signal, we aim to precisely estimate the value of an arbitrary parameter $\theta$ characterizing the scattering system ($\theta$ is here the transverse position of a target scatterer). (b)~Averaged impulse response measured when a delta pulse is generated at the input at time $t_0=0$\,\textmu s. (c)~Averaged impulse response measured for $\theta^-=\theta_0-\Delta \theta$ (blue curve) and $\theta^+=\theta_0+\Delta \theta$ (black curve), where $\Delta \theta=20$\,\textmu m. Only a short time window is represented here for the sake of clarity---this time window is represented by the black box in (b). The two curves strongly overlap, indicating that the measured output signal weakly depends on $\theta$. }
	\label{fig1}
\end{figure*}

To introduce the concepts underlying our approach, we consider a general model of time-resolved measurements performed on a given linear scattering system. This system, which can be arbitrarily complex, is parameterized by a scalar parameter $\theta$, and our goal here is to experimentally estimate its value with optimal precision from noisy measurements. This parameter can characterize any feature of the system, such as the position of a single target scatterer hidden inside a disordered medium [\fig{fig1}(a)]. To estimate the value of $\theta$, we conduct a discrete time-resolved experiment, in which an input field sampled at times $\{t_1,\dots,t_m \}$ is generated at a position $r$. The temporal distribution of the input field is conveniently described by the state $|\psi^\ins \ket$ defined in a Hilbert space of dimension $m$, equipped with the usual inner product $\bra \cdot | \cdot \ket$ and the associated Euclidean norm $\Vert {\cdot} \Vert$. In the time representation (that we note $\mathcal{T}$ representation), this state is characterized by the coefficients $\{\psi^\ins(t_0),\dots,\psi^\ins (t_{m-1})\}$. Moreover, this state is normalized so that $\Vert \psi^\ins \Vert^2=1$ (i.e. all input states have the same energy). In this way, the role of the temporal distribution of the input field (characterized by $|\psi^\ins \ket$) can be studied separately from that of its amplitude scaling factor $\mathcal{A}$. Assuming that we generate an input field $\mathcal{A} | \psi^\ins\ket$, the field propagates into the system, and an output field $\mathcal{A} | \psi^\out\ket$ sampled at times $\{t'_1,\dots,t'_n \}$ is measured at a position $r'$. As for the input field, the temporal distribution of the output field is thus described by the state $|\psi^\out \ket$, defined in a Hilbert space of dimension $n$ and characterized by the coefficients $\{\psi^\out(t'_0),\dots,\psi^\out (t'_{n-1})\}$ in the $\mathcal{T}$ representation. In the linear regime, this output state is connected to the input state through the relation $|\psi^\out\ket = U |\psi^\ins\ket$, where $U$ denotes the linear evolution operator that describes the propagation of waves into the complex scattering system, from the input emitter to the output receiver. This operator is supposed to be known, either by prior measurements or by theoretical modeling.

Because of noise fluctuations that are inherent to any measurement process, the precision at which the value of $\theta$ can be estimated from experimental data is fundamentally limited. More precisely, the variance of any unbiased estimate $\hat{\theta}$ of the parameter $\theta$ satisfies the Cram\'er-Rao inequality, which imposes that $\Ve(\hat{\theta}) \geq 1/J$ where $J$ is the Fisher information and $\Ve$ is the variance operator acting over noise fluctuations. This bound is expressed from the probability density function $p(X;\theta)$ as follows~\cite{van_trees_detection_2013}:
\begin{equation}
	J = \Ee \left(\left[ \partial_\theta p(X;\theta) \right]^2 \right) ,
	\label{definition_fisher}
\end{equation}
where $X$ is the $n$-dimensional random variable representing noisy data and $\Ee$ is the expectation operator acting over noise fluctuations. The Fisher information $J$ is by definition dependent on the observable of interest $\theta$, as well on the characteristics of the detector employed to collect the data. In principle, any noise statistics can be analyzed using this formalism. Here, we assume that any data sample $X_k$ measured at time $t'_k$ follows a Gaussian distribution of expectation value $ \mathcal{A} \psi^\out(t'_k)$ and of constant variance $\sigma^2$. This noise model not only commonly applies to measured acoustic fields, but it is also highly relevant at optical frequency (measured intensities are then typically Poissonian for shot-noise limited measurements, but complex fields recovered using e.g. an interferometric homodyne detection scheme are Gaussian when the reference beam is sufficiently strong~\cite{loudon_quantum_2000,weiner_ultrafast_2009}). Assuming that no statistical correlations exist between different sampling points and that $\sigma$ is independent of $\theta$, \eq{definition_fisher} becomes~\cite{van_trees_detection_2013}
\begin{equation}
J = \frac{\mathcal{A}^2}{\sigma^2}\sum_{k=0}^{n-1} \left[\partial_\theta \psi^\out (t'_k) \right]^2 .
\label{fisher_info}
\end{equation}
Since $|\psi^\ins\ket$ is independent of $\theta$, the derivative of the output state linearly depends on the input state through the relation $|\partial_\theta \psi^\out\ket = \partial_\theta U |\psi^\ins\ket$. Inserting this expression into \eq{fisher_info} yields the following quadratic form:
\begin{equation}
J = \frac{\mathcal{A}^2}{\sigma^2} \bra \psi^\ins |\partial_\theta U^\dagger \partial_\theta U | \psi^\ins \ket ,
\label{Fisher_general}
\end{equation}
where the symbol $\dagger$ denotes Hermitian conjugation. The operator $F= \partial_\theta U^\dagger \partial_\theta U $, which we refer to as \emph{Fisher information operator}, is Hermitian by construction. The maximum Fisher information that can be reached by shaping the input field in its temporal degrees of freedom is given by
\begin{equation}
J^\mathrm{opt} = \frac{\mathcal{A}^2}{\sigma^2} \max_{j} \Lambda_j , 
\label{eigenvalues_fisher}
\end{equation}
where $\Lambda_j$ denotes the $j$-th eigenvalue of $F$. The eigenstate $|\Phi^\ins\ket$ associated with this eigenvalue (which we call the \textit{maximum information state}) then describes the temporal distribution of the input field that must be generated to reach this optimal value. Note that, as $F$ depends on the true value of $\theta$, optimal input fields also depend on $\theta$. In practice, we will thus analyze scattering systems over restricted intervals, for which $F$ can be considered as being approximately independent on $\theta$.

The quadratic form given in \eq{Fisher_general} is valid even for time-dependent scattering media, assuming that this time-dependence is deterministic and reproducible. This general expression shows that the Fisher information carried by temporal degrees of freedom of a wave field can be quantified using a linear operator, in the exact same way as for spatial ones~\cite{bouchet_maximum_2021}. However, unlike its spatial counterpart, this operator is strongly constrained due to time translation symmetry in the common case of time-invariant scattering systems. Indeed, in this case, $U$ and $\partial_\theta U$ are both represented by Toeplitz matrices in the $\mathcal{T}$ representation~\cite{gray_toeplitz_2006}, reflecting the fact that input and output fields are simply related through a convolution operation. As a consequence, temporally-shaped maximum information states feature a number of remarkable properties that we theoretically introduce and experimentally demonstrate using acoustic waves at ultrasound frequencies.


\section{Maximizing the total Fisher information}

\label{sectIII}

In our acoustic experiment, the complex scattering system is composed of a stainless-steel waveguide (square section of $18\times18$\,mm$^2$, length $478$\,mm) immersed into water and within which thin parallel stainless-steel rods (diameter $1$\,mm) are randomly located. The parameter $\theta$ that we aim to precisely estimate is the transverse position of one of these rods [\fig{fig1}(a)], which can be deterministically and accurately translated using a motorized micropositioning stage. A wideband $5$\,MHz center-frequency transducer generates a spatially-focused ultrasound wave at the input of the waveguide, and a second transducer measures the pressure at the output of the waveguide (see Appendix~\ref{appendix_setup}). An arbitrary wavefront generator working at a repetition frequency of $1.6$\,kHz is used to generate the input state within a time window ranging from $t=0$\,\textmu s to $t=100$\,\textmu s (sampling frequency, $240$\,MHz). All input fields are generated with the same amplitude scaling factor $\mathcal{A}$; in this way, the total input energy remains constant, ensuring that all states are properly normalized in order to study only the influence of their temporal distribution. An oscilloscope then measures the output field within a time window ranging from $t'=390$\,\textmu s to $t'=490$\,\textmu s (sampling frequency, $200$\,MHz). In this way, we start the acquisition of the output field just before ballistic waves reaches the output transducer [see \fig{fig1}(b)]. Note that, in practice, input and output states are defined here from input and output electrical signals, therefore including the linear electromechanical response of the transducers.

Our approach is ultimately designed to optimally estimate small variations in the value of $\theta$ from noisy measurements. In the experiment, noise fluctuations are caused only by the additive white Gaussian noise arising from the measurement electronics, which is characterized by a standard deviation $\sigma=2.6$\,mV for single-shot measurements (see Appendix~\ref{appendix_noise}). Nevertheless, finding the maximum information state requires an accurate knowledge of the operator $\partial_\theta U$, which must be assessed as precisely as possible. In addition, the experimental characterization of the Fisher information associated with maximum information states also requires a low measurement noise. For this reason, all measurements performed during the characterization stage (Sections \ref{sectIII} and \ref{sectIV}) are averaged over $N_{\mathrm{avg}}=4096$ noise realizations. Single-shot measurements are then performed during the validation stage (Section \ref{sectV}), in order to compare the Cram\'er-Rao bound predicted from the knowledge of $\partial_\theta U$ to experimental estimations of $\theta$ carried out from noisy data.

To first experimentally assess the operators $U$ and $\partial_\theta U$, we start by placing the target at the center of the waveguide (position $\theta_0$) and we generate a delta pulse emitted at time $t=0$\,\textmu s (amplitude $10$\,V, pulse width $50$\,ns). The measured impulse response is characterized by a complicated time dependence due to the propagation of multiply-scattered waves within the disordered system [\fig{fig1}(b)]. Most importantly, this time dependence carries information about the value of $\theta$. To demonstrate this, we measure the output state for $\theta^-=\theta_0-\Delta \theta$ and for $\theta^+=\theta_0+\Delta \theta$, where $\Delta \theta=20$\,\textmu m. While similar, measured signals are not identical [\fig{fig1}(c)], indicating that the output state does (weakly) depend on $\theta$ when a delta pulse is generated at the input.

Due to time-translation symmetry, the operator $U$ is represented by a Toeplitz matrix in the $\mathcal{T}$ representation. Moreover, due to causality, this matrix is also lower triangular. Therefore, the columns of the matrices representing $U(\theta_0)$, $U(\theta^{-})$ and $U(\theta^{+})$ can simply be expressed from the impulse response measured for different values of $\theta$ (see Appendix~\ref{appendix_toeplitz}).
The centered finite-difference scheme $\partial_\theta U \simeq [U(\theta^{+})-U(\theta^{-})]/(2 \Delta \theta)$ is subsequently used to construct the derivative of the operator $U$ with respect to $\theta$. While all output signals are measured and presented at a sampling frequency of $200$\,MHz, the signal bandwidth is limited by that of the transducers (frequency bandwidth below $10$\,MHz). To avoid working with overwhelmingly large Hilbert spaces, we can therefore define input and output states at a sampling frequency of $20$\,MHz. Both $U$ and $\partial_\theta U$ are then represented by $2000\times2000$ matrices.

\begin{figure}[t]
	\begin{center}
		\includegraphics[width=8.6cm]{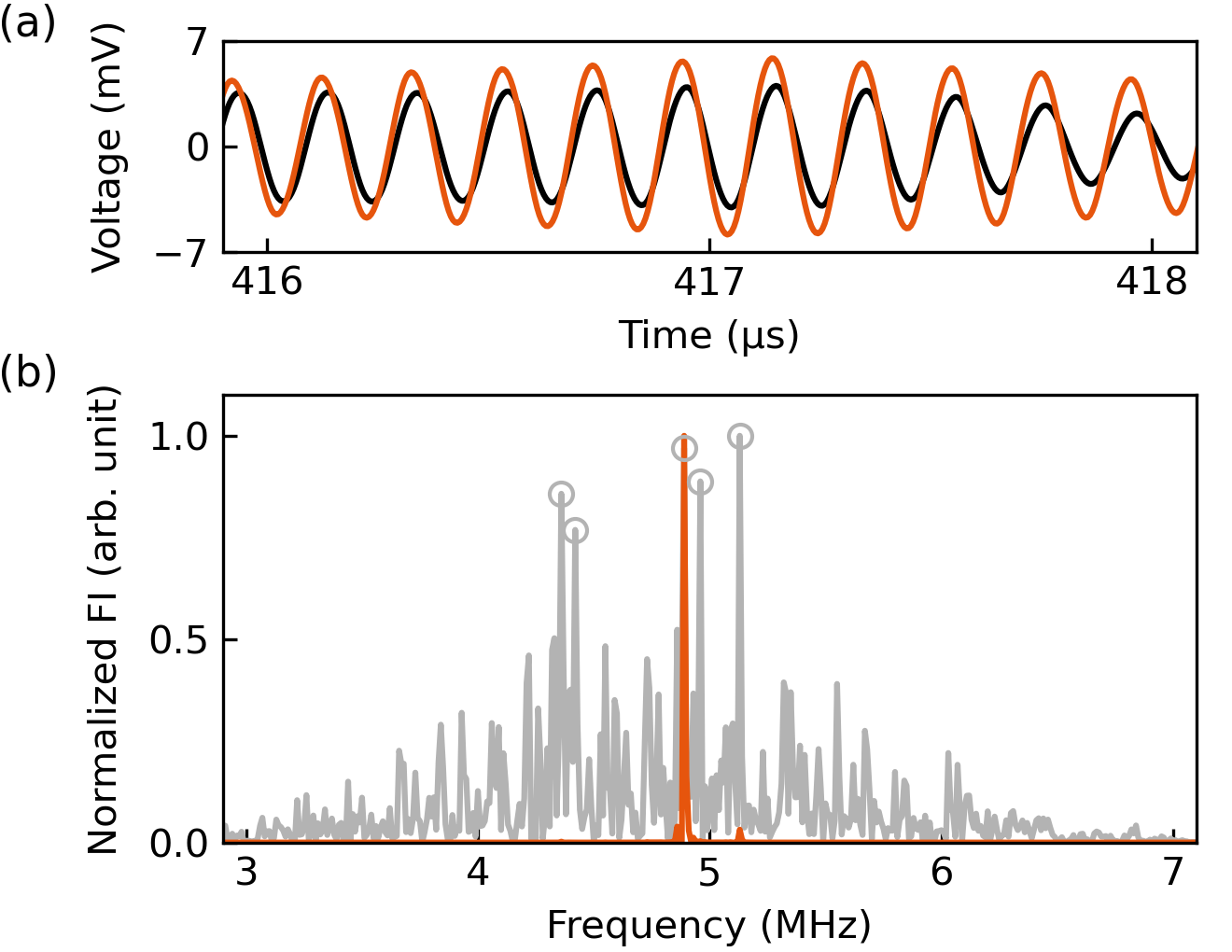}
	\end{center}
	\caption{(a) Averaged output signal measured for $\theta^-=\theta_0- \Delta \theta$ (red curve) and $\theta^+= \theta_0+\Delta \theta$ (black curve), when generating the maximum information state at the input. Only a short time window is represented here for the sake of clarity. As expected, this output signal is much more sensitive to $\theta$ as compared to the impulse response [\fig{fig1}(c)]. (b)~Normalized Fisher information per unit frequency associated with the impulse response (gray spectrum) and with the maximum information state (red spectrum). In this latter case, due to time translation symmetry, almost all the Fisher information is carried by a single frequency component ($\omega_{\mathrm{opt}}=4.89$\,MHz), which corresponds to one of the highest values of the Fisher information spectrum associated with the impulse response (for illustrative purposes, we have highlighted the $5$ highest values of this spectrum using gray circles). }
	\label{fig2}
\end{figure}

It is first instructive to study the Fisher information carried by the impulse response of the system. While this signal weakly varies with the value of $\theta$ [\fig{fig2}(a)], the achievable precision is already relatively good when all $n$ sampling points are considered altogether. Defining $|v_0\ket= \partial_\theta U | t_0 \ket$ where $| t_0 \ket$ is an input delta pulse at $t_0=0$\,\textmu s, the resulting Fisher information $J_0^{\mathrm{imp}}$ is
\begin{equation}
	J_0^{\mathrm{imp}} = \frac{\mathcal{A}^2}{\sigma^2} \Vert v_0 \Vert^2 .
\end{equation}
For a single-shot measurement in our experimental conditions ($\mathcal{A}=10$\,V and $\sigma=2.6$\,mV), we obtain $J_0^{\mathrm{imp}} =0.03$\,\textmu m$^{-2}$, which means that the Cram\'er-Rao bound on the standard error (that we call \textit{precision limit}) is $5.8$\,\textmu m. We can already remark that mapping one spatial degree of freedom (the position of the target) into many temporal ones (the time-resolved measured data) enables one to strongly surpass the resolution limit---the wavelength of ultrasound waves in water is around $300$\,\textmu m at $5$\,MHz. We can also study how the Fisher information is distributed in the frequency domain. To this end, we represent the state $|v_0\ket$ in the Fourier basis and we calculate the Fisher information associated with each frequency component. This gives us the Fisher information per unit frequency [\fig{fig2}(b), gray spectrum], which yields the total Fisher information when summed over all frequencies. This spectrum presents a number of peaks, whose distribution are seemingly random due to the presence of the disordered medium, and that look unrelated to the peaks that appear in the energy spectrum of the signal itself (see Appendix~\ref{appendix_spectra}). 

From the knowledge of the operator $F=\partial_\theta U^\dagger \partial_\theta U$, one can straightforwardly use an eigenvalue decomposition to identify the maximum information state, as indicated by \eq{eigenvalues_fisher}. We then experimentally generate this optimal input field and, as expected, we observe that the resulting output signal is much more sensitive to $\theta$ as compared to the impulse response [\fig{fig2}(a)]. This yields a total Fisher information of $0.50$\,\textmu m$^{-2}$, which means that the precision limit is $1.4$\,\textmu m, reducing the estimation error by a factor of $4$ as compared to the one obtained with the impulse response. Moreover, the Fisher information per unit frequency is sharply peaked around $\omega_{\mathrm{opt}}=4.89$\,MHz [\fig{fig2}(b), red spectrum], which corresponds to one of the highest values of the Fisher information spectrum associated with the impulse response. This observation reflects an interesting asymptotic property of maximum information states: these states are perfectly monochromatic for infinitely long time signals ($m\rightarrow \infty$ and $n\rightarrow \infty$). To demonstrate this result, we define the input field state $\psi^\ins(t)$ and the output field state $\psi^\out(t)$ using functions of $t \in ]-\infty,+\infty[$. The Fisher information $J_{\infty}$ associated with such a measurement reads
\begin{equation}
J_{\infty} = \frac{\mathcal{A}^2}{\sigma^2} \int_{- \infty}^{\infty} \left[\partial_\theta \psi^\out(t)\right]^2 \de t .
\end{equation}
Due to time translation symmetry, we can write the derivative of the output state as the convolution of the derivative $\partial_\theta u (t)$ of the impulse response $u(t)$ with the input state $\psi^\ins(t)$, which yields
\begin{equation}
\partial_\theta \psi^\out(t) = \int_{- \infty}^{\infty} \partial_\theta u(t-\tau) \psi^\ins (\tau) \de \tau .
\end{equation}
Using Parseval–Plancherel theorem along with the convolution theorem, we obtain
\begin{equation}
J_{\infty} = \frac{\mathcal{A}^2}{\sigma^2} \int_{- \infty}^{\infty} \left \vert \partial_\theta u(\omega) \right \vert^2 \left \vert \psi^\ins (\omega) \right \vert^2 \de \omega ,
\label{fisher_continuous}
\end{equation}
where $\partial_\theta u(\omega)$ and $\psi^\ins (\omega)$ are the Fourier transforms of $ \partial_\theta u(t)$ and $\psi^\ins (t)$, respectively. Then, maximizing the integral in \eq{fisher_continuous} under the normalization condition $\int_{- \infty}^{\infty} \vert \psi^\ins (\omega) \vert^2 \de \omega = 1$ amounts to find the frequency $\omega_{\mathrm{opt}}$ that maximizes the function $ \vert \partial_\theta u(\omega) \vert^2$, and to choose as an input state a monochromatic wave at $\omega=\omega_{\mathrm{opt}}$. The resulting Fisher information is
\begin{equation}
J^\mathrm{opt}_{\infty} = \frac{\mathcal{A}^2}{\sigma^2} \left \vert \partial_\theta u(\omega_{\mathrm{opt}}) \right \vert^2 .
\end{equation}
Thus, in the case of field states that are defined over $t\in]-\infty,+\infty[$, maximizing the Fisher information simply requires one to calculate the power spectrum of the function $\partial_\theta u (t) $ and to identify its maximum value. However, in our experiment, due to the finite input and output time windows over which the analysis is performed, the maximum information state is slightly different from a monochromatic field, with an envelope that is optimally tuned to account for these finite time windows (see Appendix~\ref{appendix_optimalstate}). Using a perfectly monochromatic state in our experiment would thus be sub-optimal, with a total Fisher information of $0.33$\,\textmu m$^{-2}$ instead of $0.50$\,\textmu m$^{-2}$ (i.e. a precision limit of $1.7$\,\textmu m instead of $1.4$\,\textmu m).


\section{Focusing the Fisher information in time}

\label{sectIV}

Our formalism allows one to maximize the Fisher information relative to the observable $\theta$ for any input and output time windows. As a special case, we can reduce the length of the output time window down to a single time sample $t'_l$, which will result in a direct mapping between the displacement of the target and the value of the field at this specific time. The Fisher information is then expressed by
\begin{equation}
J_l = \frac{\mathcal{A}^2}{\sigma^2}\left[\partial_\theta \psi^\out (t'_{l}) \right]^2 .
\label{single_time_fi}
\end{equation}
By maximizing $J_{l}$ over all possible input states, one can focus the Fisher information at time $t'_l$, in the same way that one can temporally focus the energy of a wave~\cite{fink_time-reversed_2000,lerosey_focusing_2007}. For this purpose, it is first required to express \eq{single_time_fi} as a function of the input state, which reads 
\begin{equation}
J_{l} = \frac{\mathcal{A}^2}{\sigma^2} \bra \psi^\ins |\partial_\theta U^\dagger |t'_{l} \ket \bra t'_{l} | \partial_\theta U | \psi^\ins \ket .
\end{equation}
The Fisher information operator that must be constructed in order to focus the Fisher information at time $t'_l$ is thus $F_l = \partial_\theta U^\dagger |t'_{l} \ket \bra t'_{l} | \partial_\theta U $. Defining $|\tilde{v}_{l} \ket = \partial_\theta U^\dagger | t'_{l} \ket$, the operator $F_{l}$ is thus simply expressed by the outer product of $|\tilde{v}_{l} \ket$ with itself, that is, $F_{l}= |\tilde{v}_{l} \ket \bra \tilde{v}_{l}| $. It can then easily be verified that all eigenvalues of $F_l$ are equal to zero, except for a single one that equals $\Vert \tilde{v}_{l} \Vert^2$. The associated eigenvector $|\Phi^\ins_l\ket = |\tilde{v}_{l} \ket/\Vert \tilde{v}_l \Vert$ is the maximum information state, and the optimal Fisher information is expressed by
\begin{equation}
J^\mathrm{foc}_{l} = \frac{\mathcal{A}^2}{\sigma^2} \Vert \tilde{v}_{l} \Vert^2 .
\end{equation}
For time-invariant systems (as in our experiment), there exists an interesting interpretation of the maximum information state $|\Phi^\ins_l\ket$ as the result of a time-reversal experiment. Assuming that $m=n$, the operator $\partial_\theta U$ is represented by a Toeplitz matrix in the $\mathcal{T}$ representation. In this case, the operators $\partial_\theta U$ and $\partial_\theta U^\dagger$ are related to each other via $\partial_\theta U^\dagger = R \partial_\theta U R $, where $R$ is the time-reversal operator which is represented by the exchange matrix in the $\mathcal{T}$ representation (see Appendix~\ref{appendix_toeplitz}). As a consequence, the state $|\tilde{v}_l\ket$ is expressed as follows:
\begin{equation}
|\tilde{v}_l \ket= R \partial_\theta U R |t'_{l} \ket .
\end{equation}
Assuming that the input and output states are sampled at the same times (up to a given time translation), we have $|t'_{l} \ket=|t_{l} \ket$. Since $R |t_{l} \ket = |t_{n-1-l} \ket $, we end up with
\begin{equation}
|\tilde{v}_{l}\ket= R \partial_\theta U |t_{n-1-l} \ket .
\label{fisher_focus}
\end{equation}
This expression demonstrates that the maximum information state $|\Phi^\ins_l\ket=|\tilde{v}_{l}\ket/ \Vert \tilde{v}_{l} \Vert$ that maximizes the Fisher information at time $t'_l$ can be interpreted as the time-reversed version of the $\theta$-derivative of the impulse response of the system when a pulse is sent at time $t_{n-1-l}$.

\begin{figure}[t]
	\begin{center}
		\includegraphics[width=8.6cm]{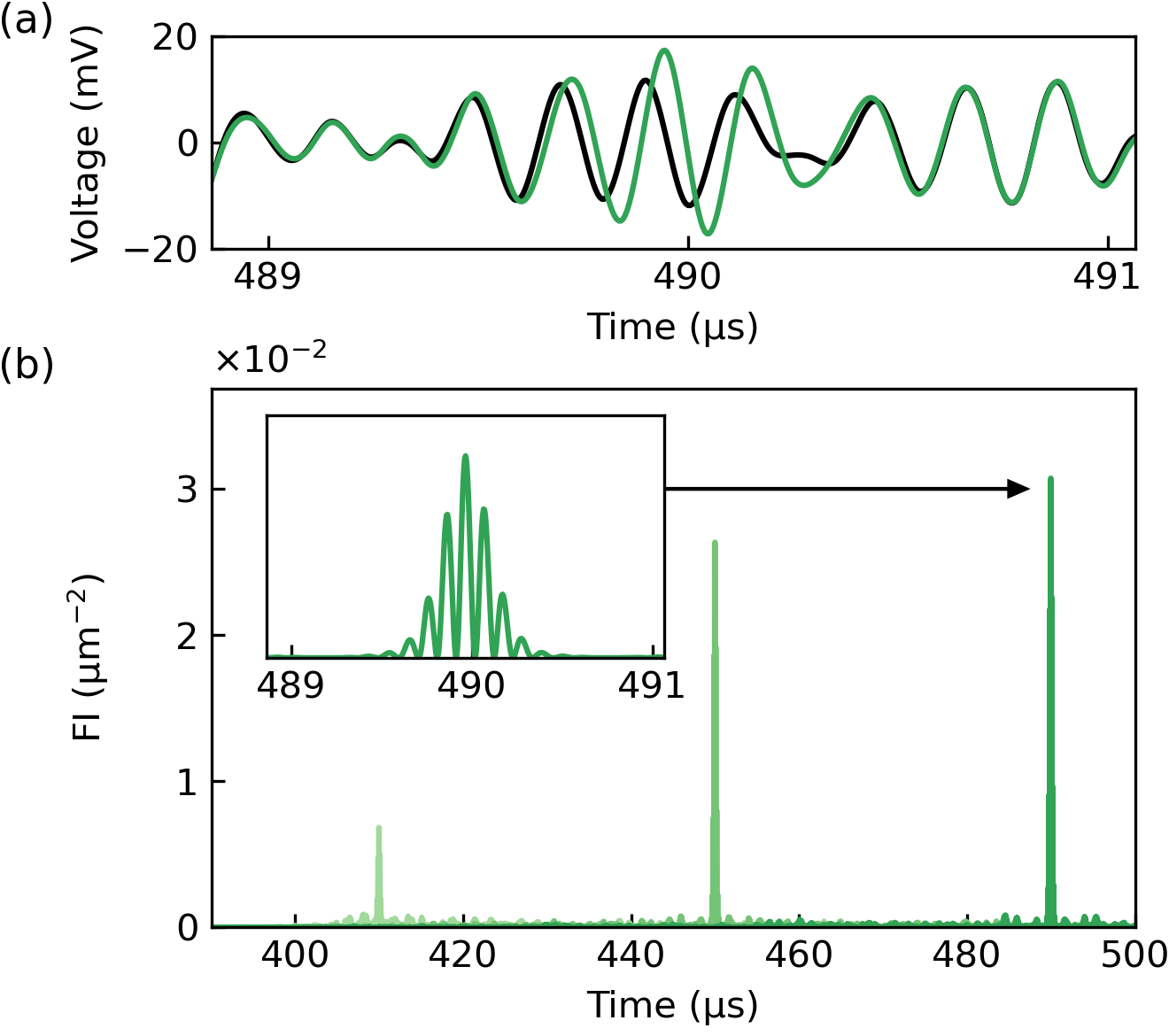}
	\end{center}
	\caption{(a)~Averaged output signal measured for $\theta^-= \theta_0- \Delta \theta$ (green curve) and $\theta^+= \theta_0+\Delta \theta$ (black curve), when generating the input state that maximizes the Fisher information at time $t'=490$\,\textmu s. Only a short time window centered around this time is represented here for the sake of clarity. The energy of the wave is not temporally focused, but the measured signal is optimally sensitive to $\theta$ at the desired time. (b)~Measured Fisher information per unit time for input states that focus the Fisher information at $t'=410$~\textmu s (light green), $t'=450$~\textmu s (medium green) and $t'=490$~\textmu s (dark green). The Fisher information is higher when the selected time is larger, since more temporal degrees of freedom of the input state can contribute to the signal. The inset provides a magnified view of the last peak, illustrating that it is composed of several lobes (these lobes are determined by the autocorrelation of the $\theta$-derivative of the impulse response).}
	\label{fig3}
\end{figure}

In our experiment, we choose to successively focus the Fisher information at three different times, namely, $t'=410$\,\textmu s, $t'=450$\,\textmu s and $t'=490$\,\textmu s. This procedure does not temporally focus the field energy but, as expected, the resulting signal is optimally sensitive to $\theta$ just at the desired time [\fig{fig3}(a)]. Representing the different output signals in the time domain and calculating the Fisher information associated with every sample point, we observe that each input field generates a peak at a different time [\fig{fig3}(b)]. These peaks have a finite width and are composed of several lobes, as determined by the autocorrelation of the $\theta$-derivative of the impulse response (see Appendix~\ref{appendix_focusstate}). Moreover, the height of these peaks increases when the Fisher information is focused at larger times, reflecting the fact that a larger number of temporal degrees of freedom of the input state can effectively contribute to the output signal while respecting causality. Interestingly, when focusing the Fisher information at time $t'_{n-1}=490$\,\textmu s, we obtain a Fisher information $J_{n-1}^{\mathrm{foc}}=0.03$\,\textmu m$^{-2}$ and a precision limit of $5.8$\,\textmu m, which are also the values obtained for the impulse response considered over the full output time windows. Indeed, the Fisher information focused at time $t'_{l}$ is, in theory, exactly equal to the Fisher information obtained for an input pulse generated at time $t_{n-1-l}$. This can be readily deduced from \eq{fisher_focus}: since $R$ is a unitary operator, we have $\Vert \tilde{v}_{l} \Vert^2 = \Vert v_{n-1-l} \Vert^2$ or, equivalently, $J_l^{\mathrm{foc}} = J_{n-1-l}^{\mathrm{imp}}$. This equivalence arises from the fact that the autocorrelation of a signal at zero delay is equal to the squared norm of this signal (see Appendix~\ref{appendix_focusstate}).


\section{Estimations from noisy data}

\label{sectV}

To demonstrate that the Cram\'er-Rao bound predicted from the measurement of the operator $\partial_\theta U$ can be reached in our experiment, we vary the position $\theta$ of the target in a step-like manner, with steps of decreasing amplitudes, and we perform a set of single-shot measurements for each position. Since we do not average over noise realizations, any given measurement $|X\ket$ is now significantly noisy (see Appendix~\ref{appendix_noise}). In order to estimate the value of $\theta$ from such noisy measurements, we employ the following linear estimator~(see Ref.~\cite{van_trees_detection_2013}, Section 5.2.4):
\begin{equation}
	\hat{\theta} -\theta_0 = \re \left( \cfrac{ \bra \partial_\theta \psi^\out | \tilde{X} - \psi^\out \ket }{ \bra \partial_\theta \psi^\out | \partial_\theta \psi^\out \ket} \right) ,
	\label{estimator_full}
\end{equation}
where $|\tilde{X}\ket = |X\ket/\mathcal{A}$ and where $| \psi^\out \ket $ is evaluated at $\theta_0$. This estimator is the minimum variance unbiased estimator and locally reaches the Cram\'er-Rao bound for Gaussian statistics. Note that, in practice, we apply this estimator from data expressed in the $\mathcal{T}$ representation, and therefore all quantities that appear in \eq{estimator_full} are real. 

\begin{figure}[t]
	\begin{center}
		\includegraphics[width=8.6cm]{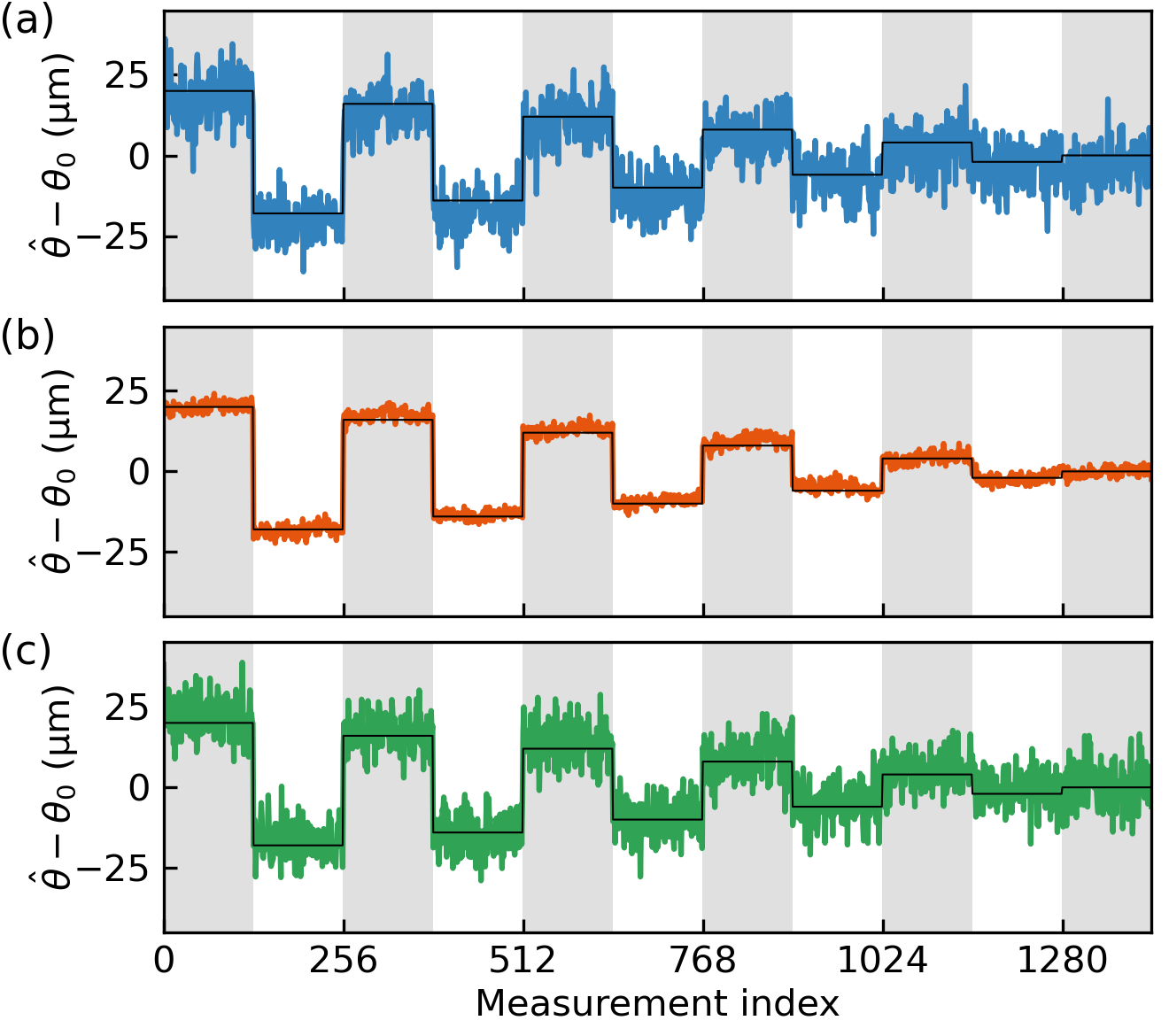}
	\end{center}
	\caption{(a) Estimated values of the displacement $\theta-\theta_0$ for a set of $1408$ successive single-shot measurements. Each measurement consists of noisy data obtained when a delta pulse is generated by the wavefront generator and when the full measured time trace (compose of $n=2000$ time samples) is processed according to \eq{estimator_full}. The observed standard error of the estimates is $6.2$\,\textmu m. (b)~Same as in (a) but for the (quasi-monochromatic) optimal input state that maximizes the Fisher information. The observed standard error of the estimates is $1.6$\,\textmu m. (c) Estimated values of $\theta$ when the input state focuses the Fisher information at time $t_{n-1}=490$\,\textmu s. For each measurement, only a single time sample is used in this case, and the measured value is processed according to \eq{estimator_partial}. The observed standard error of the estimates is $5.8$\,\textmu m. In all subfigures, the black curve represent the true value of the displacement $\theta-\theta_0$, which is deterministically controlled using a motorized stage. }
	\label{fig4}
\end{figure}

We first choose a delta pulse $|t_0\ket$ as the input state and we apply \eq{estimator_full} to the measured data set. Even though the error of the estimates is significant, the larger steps can already be distinguished [\fig{fig4}(a)]. The observed standard error of the estimates, defined as $\sigma_{\mathrm{obs}}=[\Ve(\hat{\theta}-\theta)]^{1/2}$, is $6.2$\,\textmu m, in excellent agreement with the precision limit previously predicted ($5.8$\,\textmu m). We then choose the maximum information state $|\Phi^\ins \ket$ as the input state. In this case, all steps can be much more clearly identified [\fig{fig4}(b)], and the observed standard error of the estimates reduces down to $1.6$\,\textmu m (predicted precision limit, $1.4$\,\textmu m). Finally, we choose the maximum information state $|\Phi^\ins_{n-1} \ket$, which focuses the Fisher information at time $t'_{n-1}=490$\,\textmu s. As only one time sample is involved, the linear estimator expressed by \eq{estimator_full} can be simplified into the following single-point estimator:
\begin{equation}
\hat{\theta}_l -\theta_0 = \re \left( \frac{\tilde{X}_l - \psi^\out(t'_l) }{\partial_\theta \psi^\out(t'_l)} \right) .
	\label{estimator_partial}
\end{equation}
In this case, we experimentally obtain a standard error of the estimates of $5.8$\,\textmu m (predicted precision limit, $5.8$\,\textmu m). As theoretically expected, estimated values of $\theta$ are characterized by the same variance as the ones obtained from the impulse response [Figs.~\ref{fig4}(a) and \ref{fig4}(c)], but from one single time sample instead of $n=2000$. 


\section{Conclusion}

To summarize, we demonstrated how any given parameter can be precisely estimated even in complex scattering environments by shaping the temporal degrees of freedom of an input field. For this purpose, we expressed the Fisher information associated with time-resolved measurements using a Hermitian operator. For time-invariant scattering media, this operator can be readily constructed from the impulse response of the system. We then provided an experimental validation of the approach with acoustic waves by optimally estimating small variations in the position of a scatterer inside a multiply-scattering waveguide. We experimentally demonstrated not only how to maximize the Fisher information carried by the output field, but also how to focus it at any given time. On the conceptual level, our results shed new light on time-reversal experiments: indeed, in the same way that time-reversing the field focuses the energy of the waves in time~\cite{fink_time-reversed_2000,lerosey_focusing_2007}, we evidenced that time-reversing the derivative of the field enables one to maximize the Fisher information at any given time. Our method could be generalized to multi-parameter estimations via the Fisher information matrix~\cite{van_trees_detection_2013,bouchet_optimizing_2021}, and could also be extended to classification tasks using conceptual tools such as the Chernoff bound and the Helstrom limit~\cite{cover_elements_2006,weedbrook_gaussian_2012,bouchet_optimal_2021}. By enabling a precise localization of hidden objects using acoustic waves, our approach could find interesting applications in structural health monitoring~\cite{kundu_acoustic_2014} and biomedical imaging~\cite{szabo_diagnostic_2004}. It could also be transposed to the optical regime, in order to improve the performances of techniques based on e.g. time-gated reflection
matrices~\cite{yoon_deep_2020} or time-resolved non-line-of-sight imaging~\cite{faccio_non-line--sight_2020}. Finally, our work paves the way towards a full control of the Fisher information in complex scattering systems, using temporal, spatial and quantum degrees of freedoms~\cite{pirandola_advances_2018}.


\begin{acknowledgments}
	\paragraph*{Acknowledgements.}
	The authors thank Stefan Rotter for insightful discussions, and Philippe Moreau for technical support. This work was supported by the European Research Council (ERC) within the H2020 program (grant 681514-COHERENCE).
\end{acknowledgments}

\appendix


\section{Experimental setup}

\label{appendix_setup}

\begin{figure}[ht]
	\begin{center}
		\includegraphics[width=8.cm]{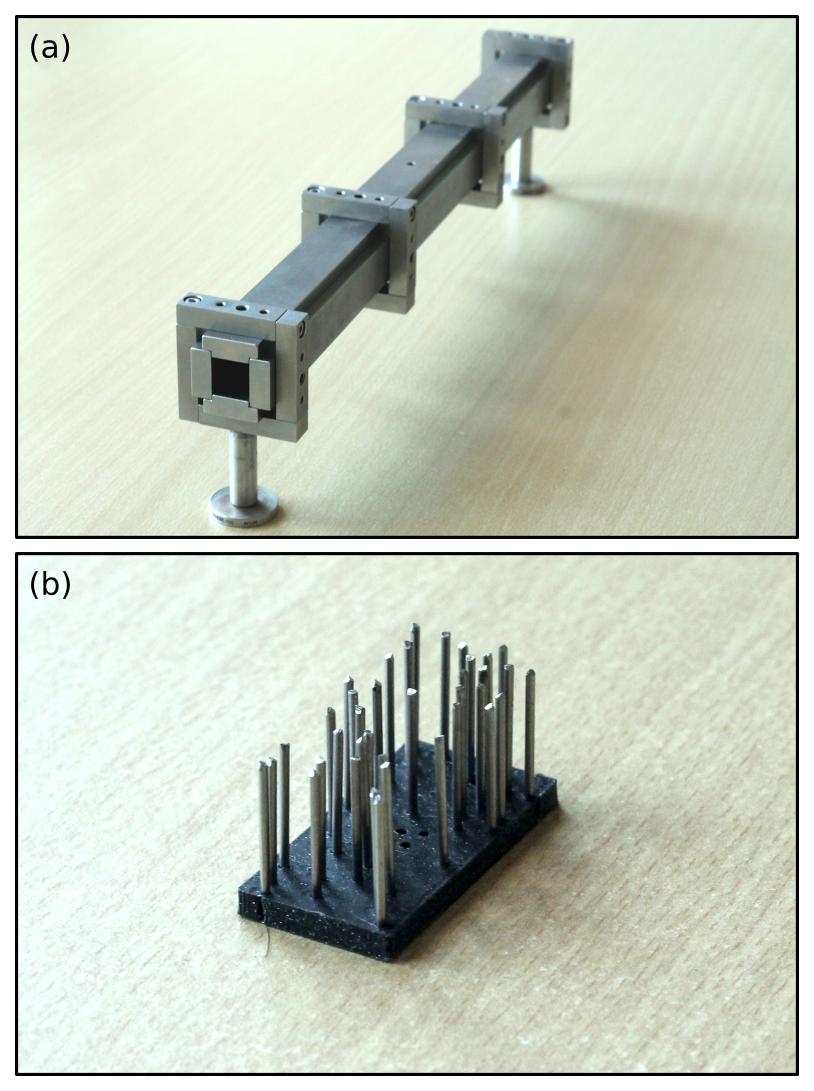}
	\end{center}
	\caption{(a)~Photograph of the stainless-steel waveguide (square section, $18\times18$\,mm$^2$ and length, $478$\,mm). (b)~Photograph of the disordered scattering medium composed of $37$ stainless-steel rods (length, $18$\,mm and diameter, $1$\,mm) that are maintained by a 3D-printed holder (black piece in the photograph, dimensions, $36$\,mm$\times 18$\,mm$\times 2.5$\,mm).}
	\label{figSI1}
\end{figure}

In our experiment, a disordered scattering medium is included into a stainless-steel waveguide characterized by a square section of $18$\,mm$\times18$\,mm and a length of $478$\,mm [\fig{figSI1}(a)]. Inside this waveguide, we place a set of $37$ stainless-steel rods of length $18$\,mm and diameter $1$\,mm [\fig{figSI1}(b)], that are randomly positioned at mid-length of the waveguide using a 3D-printed holder (dimensions, $36$\,mm$\times 18$\,mm$\times 2.5$\,mm). The waveguide is drilled at mid-length (hole diameter, 4\,mm), so that one additional stainless-steel rod (i.e. the target) can be deterministically moved from outside using a motorized staged (PI M-230.25). This system is immersed into water, within which ultrasound waves propagate efficiently. 

The input field is generated by an ultrasound transducer (Panametrics A307S, center frequency $5$\,MHz, $[3.5,6.6]$\,MHz $-6$\,dB one-way bandwidth (62$\%$), diameter $25.4$\,mm, focal length $50.8$\,mm). An arbitrary wavefront generator (Tiepie Handyscope HS5) is used to generate input signals sampled at $240$\,MHz, within a time window ranging from $t=0$\,\textmu s to $t=100$\,\textmu s, with a $14$\,bits resolution. The amplitude of any delta pulse is set to $10$\,V, with a pulse width set to $50$\,ns (this time is sufficiently brief such that the shape of the impulse response remains independent of the selected pulse width). Other temporally-shaped input signals are scaled so that their total energy remains constant.

The output field is measured using a transducer similar to the input one (Olympus V307, center frequency $5$\,MHz, $[2.5,7.2]$\,MHz $-6$\,dB one-way bandwidth (95$\%$), diameter $25.4$\,mm, focal length $76.2$\,mm). The transducer signal is first pre-amplified (Sofranel 5900 PR, gain $40$\,dB, bandpass filter $1$\,kHz/$20$\,MHz) before being digitized by an oscilloscope (Tiepie Handyscope HS5). Output signals are sampled at $200$\,MHz, within a time window ranging from $t'=390$\,\textmu s to $t'=490$\,\textmu s, with a $12$\,bits resolution. Since the same device is used as a wavefront generator and as an oscilloscope, we use its internal trigger function to synchronize emission and detection, at a repetition rate of $1.6$\,kHz. Care is taken to ensure that the amplitude fluctuations induced by the temporal jittering of the device had no significant effects throughout all measurements (i.e. for both averaged measurements and single-shot measurements). In particular, the sampling frequency of the oscilloscope ($200$\,MHz), which is much larger than the required Nyquist frequency for our signals and the bandwidth of the pre-amplifier, is chosen in order to minimize temporal jittering.


\section{Characterization of the measurement noise}

\label{appendix_noise}

\begin{figure}[ht]
	\begin{center}
		\includegraphics[width=8.6cm]{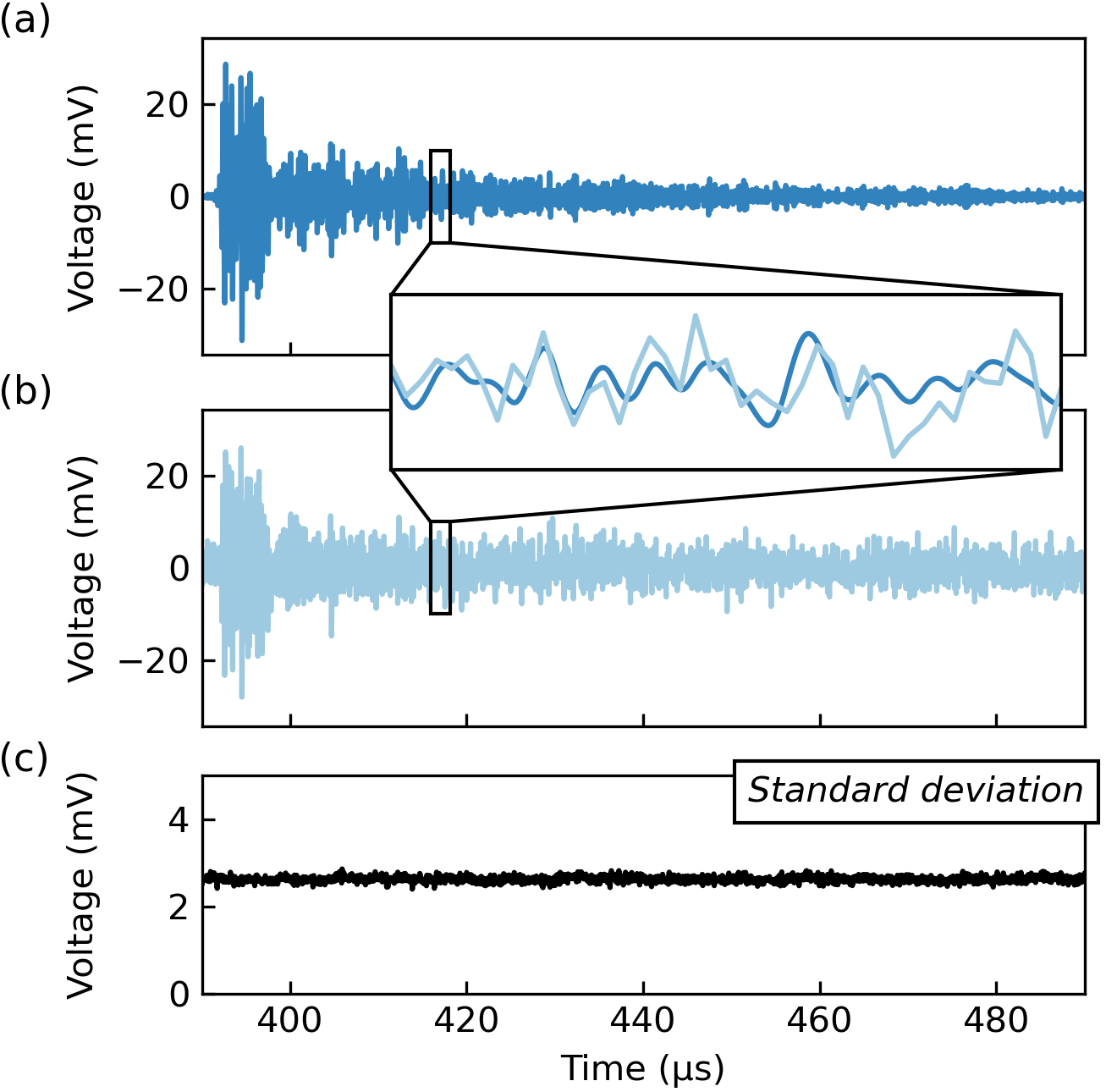}
	\end{center}
	\caption{(a)~Impulse response averaged over $N_{\mathrm{avg}}=4096$ noise realizations and smoothed using cubic splines. This constitutes a faithful estimation of the expectation value of the output field state. (b)~Single-shot measurement of the impulse response. In this case, noise significantly contributes to the measured signal. (c)~Standard deviation characterizing single-shot measurements performed with our experimental setup. The standard deviation is estimated for each time sample from an ensemble of $1024$ measurements of the impulse response. }
	\label{figSI2}
\end{figure}

To calculate the Fisher information, it is required to find a relevant model of the noise statistics. In our experiments, it was verified that in the range of measured signals (millivolt range, comparable to the standard deviation $\sigma$ of the measurement noise), the amplitude fluctuations were caused only by the additive white Gaussian noise arising from the measurement electronics, with negligible influence from any other sources of fluctuations such as the internal temporal jittering of the system. Under these conditions, any data sample $X_k$ measured at time $t'_k$ follows a Gaussian distribution of expectation value $ \mathcal{A} \psi^\out(t'_k)$ and of constant variance $\sigma^2$. In practice, two different types of measurements were performed:

\paragraph{a. Averaged measurements} To construct the operator $U$, we performed measurements averaged over $N_{\mathrm{avg}}=4096$ noise realizations.
Because measured signals were sampled at $200$\,MHz and since their actual frequency bandwidth is below $10$\,MHz, we further reduced the noise on the averaged signals by filtering out the high frequency components outside the relevant frequency band. It was checked that residual fluctuations on the final filtered averaged signals were effectively negligible, in the sense that it could be considered for our purpose that all averaged signals were measured with virtually no noise. Consequently, measured averaged signals can be considered in practice as being equal to their expectation value, as needed to construct the operators $U$ and $\partial_\theta U$. Such an averaged signal is presented in \fig{figSI2}(a).

\paragraph{b. Single-shot measurements} For single-shot measurements, noise significantly contributes to the measured signal. As introduced above, we verified that the noise could be accurately modeled by an additive white Gaussian noise. In addition, the observed standard deviation is constant over time and does not depend on the signal itself. An example of single-shot measurement is shown in \fig{figSI2}(b). The noise was accurately estimated as the standard deviation of an ensemble of $1024$ measurements, each composed of $2000$ time samples. As a result, we obtained $\sigma=2.6$\,mV [\fig{figSI2}(c)].


\section{Representation of \texorpdfstring{$\boldsymbol{\partial_\theta U}$}{the evolution operator} for time-invariant systems}

\label{appendix_toeplitz}

In order to calculate the derivative of the output state for any input state, we rely on the linear relation $|\partial_\theta \psi^\out \ket = \partial_\theta U | \psi^\ins \ket$, which is expressed as follows in the $\mathcal{T}$ representation:
\begin{equation}
\partial_\theta \psi^\out (t'_k) = \sum_{j=0}^{m-1} \partial_\theta u_{kj} \, \psi^\ins (t_j) \: .
\label{derivtive_general}
\end{equation}
This formalism is general, in the sense that it can apply to arbitrary time-dependent scattering media. Nevertheless, in our experiment, we exclusively study static scattering systems that are time invariant. As a consequence, we can write $u_{kj}=u_{k-j}$ in the $\mathcal{T}$ representation, where $u_k=u(t_k)$ can be interpreted as the discrete impulse response of the system. It follows from \eq{derivtive_general} that we can calculate the derivative of the output field using a discrete convolution operation:
\begin{equation}
\partial_\theta \psi^\out (t'_k) = \sum_{j=0}^{m-1} \partial_\theta u_{k-j} \, \psi^\ins (t_j) \: .
\label{equation_convolution}
\end{equation}
Assuming that $m=n$, the operator $\partial_\theta U$ is then represented by a Toeplitz matrix, i.e., 
\begin{equation}
\partial_\theta U = \begin{bmatrix}
\partial_\theta u_0& \partial_\theta u_{-1} & \cdots &\partial_\theta u_{-n+1} \\
\partial_\theta u_1& \partial_\theta u_0 & \ddots & \vdots \\
\vdots & \ddots & \ddots & \partial_\theta u_{-1}\\
\partial_\theta u_{n-1} & \cdots & \partial_\theta u_1& \partial_\theta u_0
\end{bmatrix}_\mathcal{T} .
\label{toepliz_matrix}
\end{equation}
In this case, the relation $|\partial_\theta \psi^\out\ket = \partial_\theta U |\psi^\ins\ket$ is simply a matrix and vector formulation of a temporal convolution of a time input with a time filter. Moreover, it is easy to verify that 
\begin{equation}
\partial_\theta U^\dagger = R \partial_\theta U R ,
\label{time_reversed_matrix}
\end{equation}
where $R$ is the time-reversal operator, which is represented by the exchange matrix
\begin{equation}
R = \begin{bmatrix}
0& \cdots & 0 & 1 \\
0 & \cdots & 1 & 0 \\
\vdots & \ddots & \vdots & \vdots\\
1 & \cdots & 0 &0
\end{bmatrix}_\mathcal{T} .
\end{equation}

In the $\mathcal{T}$ representation, the matrices that represent the operators $U$ and $\partial_\theta U$ are not only Toeplitz matrices, but they are also lower triangular due to causality. These matrices are thus not circulant matrices, and therefore they cannot be represented by diagonal matrices in the frequency representation \cite{gray_toeplitz_2006}. However, the impulse response and its derivative with respect to $\theta$ are typically localized in time (they are equal to zero before the ballistic waves reach the output transducer, and they progressively approach zero when the waves that reach the output transducer have been scattered multiples times). Thus, for long input and output time windows, the matrices representing the operators $U$ and $\partial_\theta U$ start to resemble circulant matrices, and thus their representations in the frequency domain become nearly diagonal. In this case, the Fisher information operator becomes also nearly diagonal in the frequency representation, providing us with an alternative interpretation of the fact that maximum information states are quasi-monochromatic for sufficiently-long input and output time windows.


\section{Energy and Fisher information spectra}

\label{appendix_spectra}

\begin{figure}[ht]
	\begin{center}
		\includegraphics[width=8.6cm]{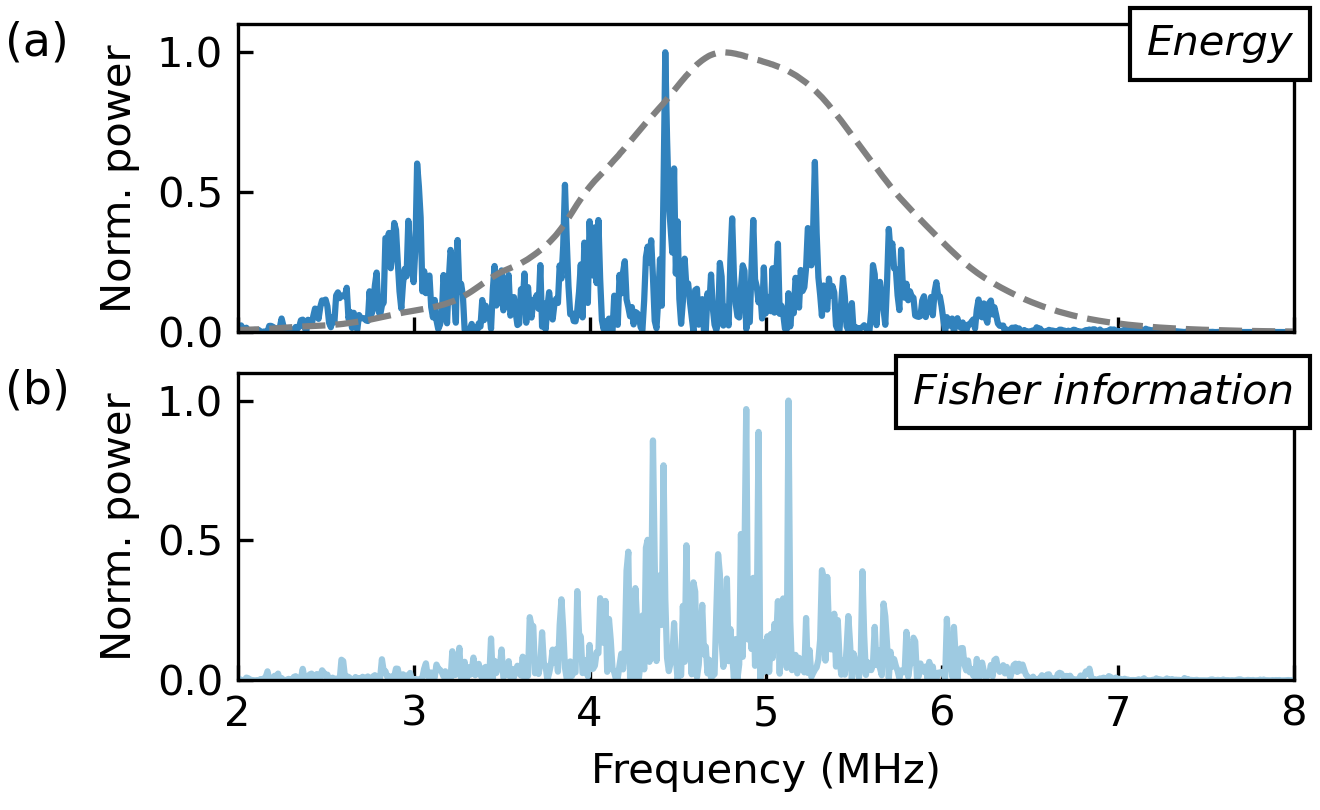}
	\end{center}
	\caption{(a)~Energy spectrum of the impulse response, when the transducers are in a confocal configuration without any scattering medium between them (dashed gray spectrum) and in the presence of the disordered waveguide (blue spectrum). (b)~Spectrum of the derivative of the impulse response, which is equivalent to the Fisher information per unit frequency. The maxima of this spectrum are different from those of the energy spectrum. }
	\label{figSI3}
\end{figure}

It is fundamentally different to study the energy of the output field and the Fisher information that it carries. As an illustration, we first show in \fig{figSI3}(a) the energy spectrum of the impulse response, when the input and output transducers are in a confocal configuration without any scattering medium between them [\fig{figSI3}(a), dashed gray spectrum]. In the presence of the complex scattering system [\fig{figSI3}(a), blue spectrum], we can see that the spectrum is strongly distorted, as expected from the disorder within the waveguide. Both of these energy spectra are also strongly different from the spectrum of the $\theta$-derivative of the impulse response (i.e. the Fisher information spectrum), which is represented in \fig{figSI3}(b). In particular, the position of the maxima are strongly different, which implies that maximizing the transmitted energy and the delivered Fisher information involve different input fields. Notably, in our experiment, the input state that maximizes the total energy of the output state (calculated from an eigenvalue decomposition of $U^\dagger U$) is a quasi-monochromatic field at $4.43$\,MHz. In contrast, the maximum information state is a quasi-monochromatic field at $4.89$\,MHz.

\section{Maximum information state defined over the full time windows}

\label{appendix_optimalstate}

\begin{figure}[ht]
	\begin{center}
		\includegraphics[width=8.6cm]{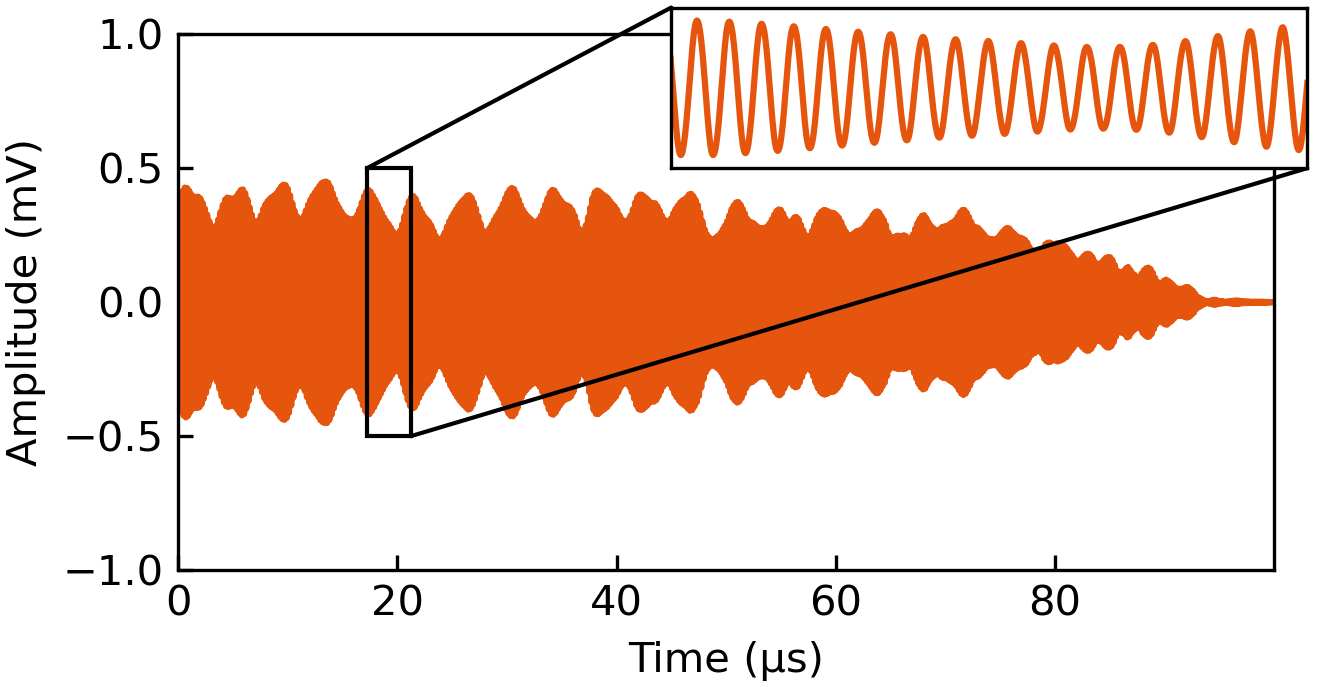}
	\end{center}
	\caption{Temporal dependence of the maximum information state, identified using an eigenvalue decomposition of the operator $F=\partial_\theta U^\dagger \partial_\theta U$. This field is quasi-monochromatic but presents an envelope that optimally takes into accounts for the finite input and output time windows. }
	\label{figSI4}
\end{figure}

For fields defined over infinitely-long time windows, the maximum information state is a monochromatic field at the frequency that maximizes the $\theta$-derivative of the impulse response of the system. However, in our experiment, we work with finite time windows, ranging from $t=0$\,\textmu s to $t=100$\,\textmu s at the input and from $t'=390$\,\textmu s to $t'=490$\,\textmu s at the output. For this reason, the maximum information state, calculated from an eigenvalue decomposition of the operator $F=\partial_\theta U^\dagger \partial_\theta U$, is not perfectly monochromatic, but presents an envelope that optimally accounts for these finite time windows (\fig{figSI4}). We notably observe that the end of the input signal is dampened. Indeed, due to causality, most of the waves that are generated close to the end of the input window will not reach the output transducer before $t'=490$\,\textmu s, and therefore do not contribute to the total Fisher information calculated up to $t'=490$\,\textmu s. This explains why the maximum information state presents an envelope that decreases over time, at the cost of introducing additional frequencies into its spectrum. 

\section{Temporal dependence of the focused Fisher information}

\label{appendix_focusstate}

The Fisher information associated with any output time sample $t'_k$ is expressed as follows:
\begin{equation}
	J(t'_k) = \frac{\mathcal{A}^2}{\sigma^2} \left[ \partial_\theta \psi^\out(t'_k) \right]^2 .
	\label{fi_single_time_1}
\end{equation}
In the case of time-invariant systems, we can use \eq{equation_convolution} to express the derivative of the output state. Then, \eq{fi_single_time_1} becomes
\begin{equation}
J(t'_k) = \frac{\mathcal{A}^2}{\sigma^2} \left( \sum_{j=0}^{m-1} \partial_\theta u_{k-j} \, \psi^\ins (t_{j}) \right)^2 .
\end{equation}
Even though $\partial_\theta u_j$ and $\psi^\ins (t_{j})$ were originally defined only for $j \in \{ 0,\dots,m-1 \}$, we can extend their definition to $j \in \mathbb{Z}$ by assuming that they are equal to zero if $j<0$ or if $j>m-1$. This procedure yields
\begin{equation}
J(t'_k) = \frac{\mathcal{A}^2}{\sigma^2} \left( \sum_{j=-\infty}^{\infty} \partial_\theta u_{j} \, \psi^\ins (t_{k-j}) \right)^2 .
\end{equation}
In order to maximize the Fisher information at time $t_{n-1}$ (last sample point of the output window), the input state must be $|\Phi^\ins_{n-1}\ket = |\tilde{v}_{n-1}\ket/\Vert \tilde{v}_{n-1} \Vert = \partial_\theta U^\dagger |t'_{n-1}\ket/\Vert \tilde{v}_{n-1}\Vert$ (as demonstrated in the manuscript for any sample point at $t'_l$). This yields
\begin{equation}
J(t'_k) = \frac{\mathcal{A}^2}{\sigma^2 \Vert \tilde{v}_{n-1} \Vert^2} \left( \sum_{j=-\infty}^{\infty} \partial_\theta u_{j} \, \partial_\theta u_{j+n-1-k} \right)^2 .
\label{eq_autocorrelation}
\end{equation}
\Eq{eq_autocorrelation} shows that the temporal dependence of the Fisher information is exactly the square of the autocorrelation of the $\theta$-derivative of the impulse response, centered at $t'_{n-1}$. 
Evaluating this expression at $t'_{n-1}$ yields 
\begin{equation}
J(t'_{n-1}) = \frac{\mathcal{A}^2}{\sigma^2 }  \sum_{j=-\infty}^{\infty} (\partial_\theta u_{j})^2 .
\label{eq_autocorrelation_zerodelay}
\end{equation}
Here, we recover the fact that focusing the Fisher information at a given time results in a Fisher information that is equal to the Fisher information enclosed in the full impulse response of the system.

\begin{figure}[ht]
	\begin{center}
		\includegraphics[width=8.6cm]{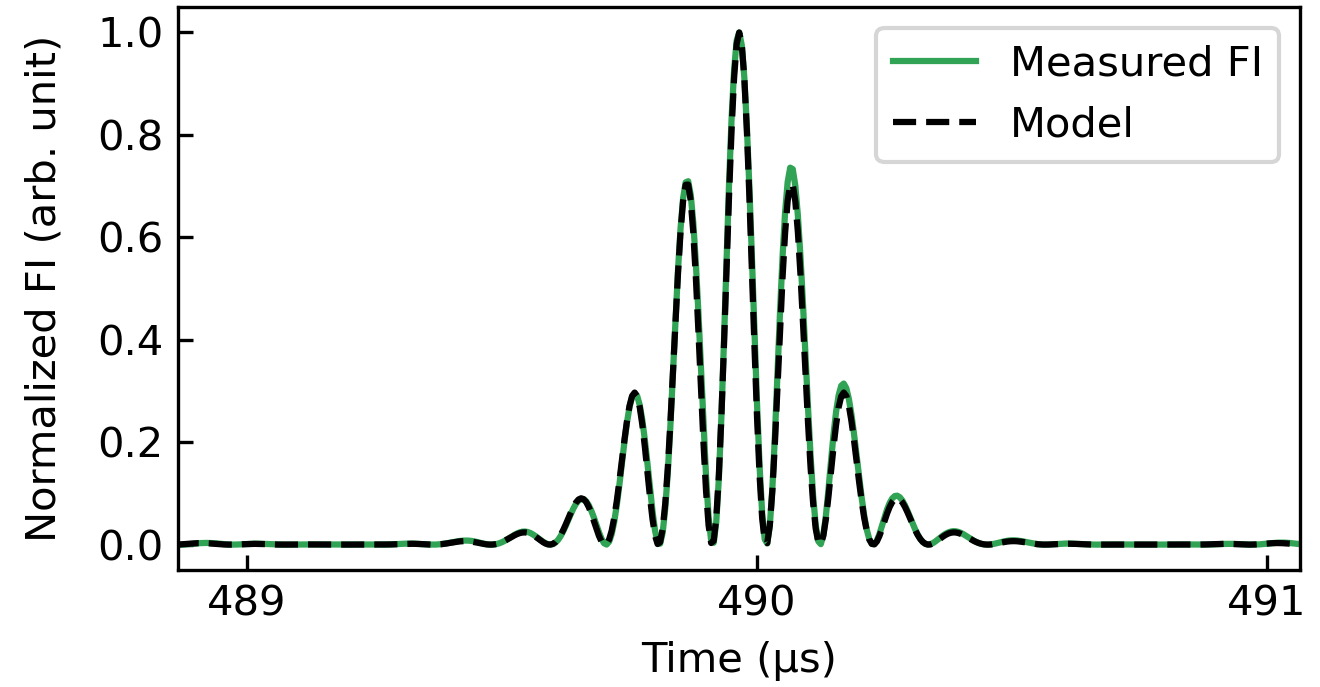}
	\end{center}
	\caption{Fisher information per unit time measured when generating the maximum information state $|\Phi^\ins_{n-1}\ket$ at the input (green curve), compared to the autocorrelation of the $\theta$-derivative of the impulse response centered at time $t'_{n-1}$ (dashed black curve).}
	\label{figSI5}
\end{figure}

In order to illustrate experimentally the property expressed by \eq{eq_autocorrelation}, we show in \fig{figSI5} the Fisher information per unit time that we measured when using the maximum information state $|\Phi^\ins_{n-1}\ket$ as an input state (green curve), as well as the autocorrelation of the $\theta$-derivative of the impulse response (dashed black curve) centered at $t'_{n-1}$. Both curves are in excellent agreement, which confirms our theoretical predictions expressed by \eq{eq_autocorrelation}.

%
	
\end{document}